\documentclass{iopart}
\usepackage{iopams}  
\usepackage[colorlinks]{hyperref} %links for citations
\usepackage{graphicx}% Include figure files
\usepackage[dvips]{color}
\usepackage{amssymb}
\usepackage{amsbsy}
\usepackage{amsfonts}
\usepackage{mathptm}
\usepackage{wasysym}
\usepackage{mathptm,times}
\usepackage{anysize,color}
\usepackage[colorlinks]{hyperref} 
\usepackage[dvips]{color}
\usepackage{graphicx}
\usepackage{bm}

\newcommand{\bra}[1]{\langle #1|}
\newcommand{\ket}[1]{|#1\rangle}

\begin{document}

\title{Non-Abelian anyonic interferometry with a multi-photon spin lattice simulator}
\author{D. W. Berry$^{1,2}$,\ M. Aguado$^3$,\ A. Gilchrist$^2$,\ and G.K. Brennen$^2$}
\address{$^1$ Institute for Quantum Computing \& Department of Physics and Astronomy, 
University of Waterloo, 200 University Avenue West, Waterloo, Ontario N2L3G1, Canada \\ $^2$ QSciTech \& Department of Physics, Macquarie University, 2109, NSW Australia\\
$^{3}$Max-Planck-Institut f\"ur Quantenoptik, Hans-Kopfermann-Str. 1, D-85748 Garching Germany}

\date{\today}
\begin{abstract}
Recently a pair of experiments \cite{Lu, Pachos:07} demonstrated a simulation
of Abelian anyons in a spin network of single photons.  The experiments were
based on an Abelian discrete gauge theory spin lattice model of Kitaev
\cite{Kitaev:03}.  Here we describe how to use linear optics and single photons
to simulate non-Abelian anyons.  The scheme makes use of joint qutrit-qubit
encoding of the spins and the resources required are three pairs of parametric
down converted photons and $14$ beam splitters.  
\end{abstract}
\maketitle

\section{Introduction}
%A compelling argument for pursuing quantum technologies is to better understand our natural world.
We know that even when the local physics of a system is understood the global behavior can yield surprising features that are difficult to predict. Well known examples are critical phenomena
and symmetry breaking. Topologically ordered media are another example, where one can have a many body system with a strongly correlated ground ``vacuum'' state with more symmetry that the microscopic equations of motion themselves. This is manifested in two dimensional electron gases  in the quantum Hall effect, where at fractional filling of the particle orbitals, the low lying excitations above the ground state are particle-like and predicted to behave like anyons \cite{Arovas:84,Moore:91, Nayak:96}.
 As point-like particles anyons exist only in two dimensions, and unlike Bosons or Fermions which have a $\pm 1$ phase under exchange, they accumulate an arbitrary phase in the Abelian case, or a matrix valued action on the Hilbert space of particle fusion outcomes in the non-Abelian case \cite{Leinaas,Goldin:85}.

To date there has been some experimental evidence in support of the existence of Abelian anyons in fractional quantum Hall fluids \cite{Camino:05a, Camino:05b} though there are caveats to interpretation of the data \cite{Fiete:07}, and a vigorous experimental effort is in place to demonstrate interferometry on non-Abelian anyons.  Recently two experiments have reported observation of $e/4$ electron charge in the $\nu=5/2$ filled state \cite{Dolev:08,Willett:09} which is predicted to have non-Abelian quasiparticles.  A theoretical interpretation of the data from Ref. \cite{Willett:09} and suggestions for a witness to non-Abelian statistics in that experimental setup is given in \cite{Bishara:09}.  There are several proposals for performing non-Abelian interferometry in the $\nu=5/2$ filled quantum Hall state \cite{Fradkin:98,DasSarma:05,Stern:06}, and for the general theory of interferometry of non-Abelian anyons see Refs. \cite{Bais:01,Bonderson:06a,Bonderson:06b}. 

A promising alternative platform to observe anyons is in spin lattices.  Here the strategy is to take a two dimensional array of spins (i.e. a group of spins whose coupling graph defines a surface) and prepare a highly entangled ground vacuum state with respect to some physical theory.  Early work by Bais \cite{Bais:80} showed that discrete gauge theories in two dimensions could support non-Abelian particles and the full mechanics of braiding in those models was given in \cite{Propitius:95}.  Kitaev \cite{Kitaev:03} suggested a spin lattice Hamiltonian that realizes discrete gauge theory on a surface with ground states that are invariant under local gauge transformations generated by elements of a discrete group.  In the case of the simplest Abelian group $G=\mathbb{Z}_2$, the microscopic spins are qubits and the model is known as the surface code Hamiltonian:
\[
H=-\sum_v A(v) -\sum_f B(f)
\]
where spins are placed on the edges of a lattice.  Here interactions amoung edges meeting at a vertex $v$ [a neighborhood denoted ${\rm star}(v)$] are given by the vertex operators
$A(v)=\prod_{\rm e\in {\rm star}(v)}\sigma_e^x$ while the interactions on the edges on the boundary of a face $f$ (a neighborhood denoted $\partial f$) are given by the face operators $B(f)=\prod_{e\in \partial f}\sigma_e^z$.  The face and vertex terms can be thought of as projections onto gauge invariant states.  The ground state has $+1$ outcome for measurement of either a face or vertex operator, and a $-1$ outcome of measuring $A(v)$ or $B(f)$ is equivalent to a particle residing on a vertex or face of the lattice; i.e. the particle states are eigenstates of the Hamiltonian.

Several proposals have been made for engineering the surface code Hamiltonian in physical systems, and to demonstrate interferometry of Abelian anyons (for a review see \cite{BrennenPachos} and references therein). These proposals suggest realizing the Hamiltonian in the perturbative regime corresponding to the gapped phase of native two body Hamiltonian on a honeycomb lattice that is realistic \cite{Kitaev:06}, and can be designed in the laboratory.  With an additional magnetic field perturbation, the honeycomb model also supports a phase with non-Abelian anyons and there are proposals for observing the statistics in these models \cite{Lahtinen:09,Yu:08}.  Such simulations require sufficiently large lattices to overcome finite size effects \cite{Kells:08}.

Another approach is to not try to build the Hamiltonian at all but rather just simulate the kinematics of the anyonic states.  That would involve creating the highly entangled ground state, and performing operations on the spins for creating, braiding and annihilating excitations.  The operations are the same regardless of whether the background Hamiltonian is present.  Such a \emph{digital simulation} of the model is not topologically protected since there is no gap to creating excitations, but all the same measurement outcomes are obtained provided the operations are done quickly relative to decoherence time scales. 
A promising platform for this kind of digital simulation is to use entangled states of single photons. This brings the advantage of easy single qudit manipulation with passive optical elements, low decoherence, and well characterised loss channels. Entangling gates between qudits are more difficult, but there are well studied pathways for nondeterministically creating such operations. The key limitations for optics are a poor single photon sources, limiting experiments to about six photons, and the difficulty of building circuits with large depth. Large depth amounts to complex nested interferometers which require tuning and stabilisation to a fraction of the wavelength. Recent advances in integrated quantum photonics \cite{Politi:08, Marshall:09} hold the key to increasing the depth of circuits that can be created, and we envisage using just such a platform for simulating non-Abelian anyons.    

\begin{figure}[!thb]
\center{\includegraphics[scale=0.7]{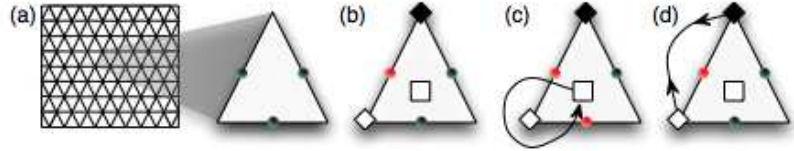}}
\caption{A simulation of non-Abelian anyonic interferometry using entangled photons (based on a generic protocol in \cite{Aguado:08a,Brennen:09}).  (a)  Begin with a surface cellulation where multilevel spins reside on all the links.  We could imagine that there is a background topological Hamiltonian, $H_{\rm TO}$, and the spins are prepared in a highly entangled state corresponding to the ground state of $H_{\rm TO}$. (b) A single plaquette of the lattice where an operation has been done on one spin, indicated in red, which creates localized anionic excitations in $H_{\rm TO}$. The electric charge/anti-charge excitations are drawn as diamonds, the flux excitations as squares. (c) Braiding of the flux around one charge. (d)  Fusion of the electric charges to measure the effect of non-Abelian braiding. An implementation with non-interacting photons would have $H_{\rm TO}=0$, yet provided the entangled vacuum state is prepared and the same operations are done as if $H_{\rm TO}$ were present, one obtains the same measurement outcomes.}
\label{fig:1}
\end{figure}

It is challenging to build up many body entanglement using single photons and linear optics, yet even the small size scale properties are rich.  In fact, it is the very virtue of topologically ordered systems that the physics is scale invariant.   As shown in Fig.~\ref{fig:1} a single plaquette of a spin lattice is enough to demonstrate braiding statistics of non-Abelian anyons.  Of course for such a small system it would be easy to perform a classical simulation to predict such behavior.  Yet the point would be to exhibit the emergent behavior on small prototypes using tools of coherent control that could be extended to scalable systems.    Two recent experiments have demonstrated a simulation of Abelian statistics in a single plaquette of a spin lattice \cite{Lu,Pachos:07} using six and four photons respectively.  A simulation using NMR control of $4$ carbon atoms in a molecule of crotonic acid was also realized \cite{Du:07}.   A simulation of non-Abelian statistics would mark significant progress for two reasons:  first, the effect of non-Abelian statistics is more dramatic and easier to discriminate from geometric or dynamical phases, and second, several novel phenomena occur such as the existence of topological entropy which could, in principle, be measured in small scale systems.     The trade off is an increased complexity in the physical operations.  Yet for our scheme only three type-I parametric down conversion crystals are needed to prepare the initial state, which is the same number used in the Lu {\it et al.} \cite{Lu} experiment to measurement Abelian statistics.

The manuscript is organized in three stages: the generic lattice model; the mapping to a single plaquette system; and the mapping to a physical system.  Readers just wishing to see the details of the steps involved in the experimental proposal can skip to Sec.~\ref{sec:4}.  In Sec.~\ref{sec:2} we introduce the simplest spin lattice model of the discrete gauge theories of Kitaev that has non-Abelian anyonic excitations.  There are several different types of non-Abelian particles arising from this model and in Sec.~\ref{sec:3} we describe one of several types of braiding experiments that could be done to reveal the non-Abelian statistics.  The operations are done using six level particles which can be encoded in a qutrit-qubit pair.  A physical realization of these ideas using linear optical elements and single photons is given in Sec.~\ref{sec:4}.   Finally we conclude in Sec.~\ref{sec:5} with a discussion of the viability of this technique and possible further avenues for experimental inquiry.

\section{Spin lattice model for non-Abelian anyons}
\label{sec:2}
\subsection{Non-Abelian discrete gauge theory}
Kitaev's generalization of the surface code Hamiltonian is also a sum of vertex and face operators and has localized particle-like excitations, but now the ground states are invariant under gauge transformations generated by some finite group $G=\{g_j\}$ of choice.  We consider the smallest non-Abelian 
group $G=S_3$, which is the permutation group on three objects.
To specify the spin lattice we consider a celluation of a two dimensional surface with the vertex set $\mathcal{V}=\{v_i\}$, edge
set $\mathcal{E}=\{e_j\}$, and face set $\mathcal{F}=\{f_j\}$.    Qudits with $d=|G|=6$ levels are placed on the edges and physical states are elements of a Hilbert space $\mathcal{H}=\mathcal{H}(6)^{\otimes |\mathcal{E}|}$ where 
$\mathcal{H}(6)=\mathbb{C}\ket{0}+\cdots +\mathbb{C}\ket{5}$.  Particles on edges that meet at a vertex $v$ all interact via a vertex operator $A(v)$.   Similarly, all particles on edges that are on the boundary of a face $f$ interact via $B(f)$.  We pick an orientation for each edge with $e=[v_j,v_k]$
denoting an edge with arrow pointing from vertex $v_j$ to $v_k$.  The choice of edge orientations is not important as long as a consistent convention is used.  We assume an orientable complex $\Gamma$ and each face $f$ has an orientation consistent with it.  The Hamiltonian is a sum of operators chosen such that the ground states of $H_{\rm TO}$ are invariant under local gauge transformations
\begin{equation}
T_g(v)=\prod_{e_j\in [v,*]}L_g(e_j)\prod_{e_j\in [*,v]}R_{g^{-1}}(e_j),
\end{equation}
Here $L_g(e_j),R_g(e_j)\in U(6)$, the $6$ dimensional unitary group, are the permutation representations of the left and right action of multiplication by the group element $g\in S_3$ on the system particle located at edge $e_j$. For the particle states we make the identification $\ket{j}\equiv\ket{g_j}$, where by convention $\ket{0}\equiv\ket{g_0}\equiv\ket{e}$, with $e$ the identity element. The action of left and right group multiplication on the basis states is then $L_h\ket{j}=\ket{h g_j}$, and $R_h\ket{j}=\ket{g_j h}$.

The spin lattice model is:
\begin{equation}
H_{\rm TO}=-\sum_v A(v)-\sum_f B(f)
\end{equation}
where 
\begin{equation}
\begin{array}{lll}
A(v)&=&\frac{1}{6}\sum_{g\in S_3}T_g(v), \\
B(f)&=&\displaystyle{\sum_{\{h_k | \prod_{e_k\in\partial f}h_k=e\}}\bigotimes_{k | e_k\in \partial f}\ket{h_k^{-o_f(e_k)}}_{e_k}\bra{h_k^{-o_f(e_k)}}}
\end{array}
\end{equation}

\begin{figure}[thb]
\center{\includegraphics[scale=0.8]{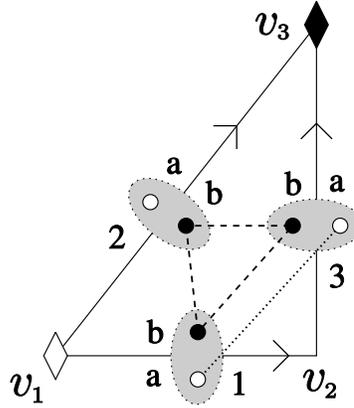}}
\caption{The triangle of nodes on which to demonstrate the non-Abelian topological action. The white(black) diamonds indicate the vertices where the charges(anti-charges) reside, and the arrows indicate the directions of the edges.    The dotted ellipses indicate the 6-level qudits on the edges. Each qudit consists of a qutrit, labelled $b$ and indicated by the solid circle, and a qubit, labelled $a$ and indicated by the empty circle. The dashed lines indicate the entangled state that needs to be created between the three qutrits, and the dotted line indicates the two qubits that must also be entangled.}
\label{fig:tri}
\end{figure}

In the definition of $B(f)$, the sum is taken over all products of group elements $h_k$ acting on  a counterclockwise cycle of edges on the boundary of $f$ such that the accumulated left action is the identity element $e\in S_3$ (i.e. $h_{\ell}h_{\ell-1}\ldots h_2h_1=e$ for the counterclockwise cycle starting at edge $e_1$ and ending at edge $e_{\ell}$).  The function $o_f(e_j)=\pm 1$ according to whether the orientation of the edge is the same as (or opposite to) the face orientation.
By construction $[A(v),A(v')]=[B(f),B(f')]=[A(v),B(f)]=0$.  Furthermore, it is straightforward to verify that since $A(v)$ is a symmetrized gauge transformation it is a projection, as is $B(f)$.  The ground states of $H_{\rm TO}$ are then manifestly gauge invariant states.  For a planar surface with boundary, the ground state $\ket{GS}$ of $H_{\rm TO}$ is unique. 
Excited states can be described by localized particles in the sense that the expectation value $\langle A(v)\rangle =0$ corresponds to an excited state with an electric charge located at vertex $v$ while  $\langle B(f)\rangle =0$ indicates a magnetic flux at face $f$.  These particles have anyonic statistics and we refer the reader to Refs. \cite{Aguado:08a,Brennen:09} for a detailed description of the particle spectrum of this model and the operations used to create, manipulate and fuse the anyons.
% corresponding to the irreps of $D(S_3)$ labeled by $\Pi^{[\alpha]}_{R(N_{[\alpha]})}$ where 
%$[\alpha]$ denotes a conjugacy class of $G$ which labels the magnetic charge, and $R(N_{[\alpha]})$ denotes a unitary irrep $R$ of the centralizer of an element in the conjugacy class $[\alpha]$ which labels the electric charge.  Pure electric charges(fluxes) represent violations of vertex(face) operators and are said to reside on vertices(faces) of the lattice.  

The specific interferometry experiment we consider is the creation of a fluxless electric charge pair, followed by the braiding of a magnetic flux around one charge, and the subsequent fusion of the charges.  This entire process can be simulated on a single lattice plaquette as summarized in Fig.~\ref{fig:1}.  Pure electric charges are labeled by irreps $R$ of the group $S_3=\{e,c_+,c_-,t_0,t_1,t_2\}$, where $e$ is the identity, $t_j$ are transpositions, and $c_{\pm}$ are $3$-cycles.  The non-Abelian electric charges are contained in the two dimensional irrep of $S_3$:
\begin{equation}
R_2 ( e )
=
\mathbf{1}_2 ,\quad
R_2 ( t_k )
=
\sigma^x e^{
      i\, \frac{ 2 \pi }{ 3 } \, k \, \sigma^z
} ,\quad 
R_2 ( c_{\pm} )
=
e^{
    \pm  i\, \, \frac{ 2 \pi }{ 3 } \, \sigma^z
} .
\label{2dirrep}
\end{equation}
%   All pure electric charge pair states satisfy $A(v'')\ket{{\bf 1}_{R};(v,v')}=(1-\delta_{v,v''})(1-\delta_{v'\!,v''}) \ket{{\bf 1}_{R};(v,v')}$ and $B(f)\ket{{\bf 1}_{R};(v,v')}=\ket{{\bf 1}_{R};(v,v')} \forall f$. Here and throughout, we consider pure electric charges so that $R$ is an irrep of the centralizer of the identity element, i.e. $R$ is an irrep of the group $S_3$ itself. 
Consider the lattice with a single face and three edges indicated in Fig.~\ref{fig:tri}.  The edges are oriented $e_1=[v_1,v_2],e_2=[v_1,v_3],e_3=[v_2,v_3]$.  For simplicity of notation we label the edges by the numbers $1,2,3$.  The ground state is quite easy to write down:
\begin{equation}
\ket{GS}=\frac{1}{6}\sum_{g,k\in S_3}\ket{k}_1\otimes\ket{g}_2\otimes\ket{k^{-1}g}_3=\frac{1}{6}\sum_{g,k}\ket{g}_2\ket{k,k^{-1}g}_{1,3}
\end{equation}
The state of a generic electric charge anti-charge pair located at vertices $(v_1,v_3)$ in the irrep $R$ (with dimension $|R|$) is given by
\begin{equation}
| M_R; ( v_1, v_3 ) \rangle
\equiv
\frac{ 1 }{ 6}
\sum_{g\in S_3} \mathrm{tr} \{ M_R R^{\dagger} (g) \} | g \rangle_2
\sum_{k\in S_3} | k, \, k^{-1} g \rangle_{1,3}
\end{equation}
with $M_R$ an $|R| \times |R|$ matrix normalized so that $\sum_{a,b=1}^{|R|}|(M_R)_{a,b}|^2=|R|$.  Under a local gauge transformation $T_h$ at vertex $v_1$, the basis states undergo the mapping:  $T_h(v_1)\ket{g}_2\ket{k,k^{-1}g}_{1,3}=\ket{hg}_2\ket{hk,k^{-1}g}_{1,3}$, and the change in the charge state is:
\begin{equation}
\begin{array}{lll}
T_h(v_1)| M_R; ( v_1, v_3 ) \rangle
&=&
\frac{ 1 }{ 6 }
\sum_{g,k\in S_3}\mathrm{tr} \{ M_R R^{\dagger} (g) \}
| hg \rangle_2| hk, \, k^{-1} g \rangle_{1,3}\\
&=&
\frac{ 1 }{ 6 }
\sum_{g'\!,k\in S_3}\mathrm{tr} \{ M_R R^{\dagger} (h^{-1}g') \}
| g' \rangle_2| hk, \, k^{-1} g \rangle_{1,3}\\
&=&
\frac{ 1 }{ 6 }
\sum_{g'\!,k'\in S_3}\mathrm{tr} \{ M_R R^{\dagger}(g')R(h) \}
| g' \rangle_2 | k'\!, \, k'^{-1} g' \rangle_{1,3}\\
&=&| R(h)M_R; ( v_1, v_3 ) \rangle.
\end{array}
\end{equation} 
where we have used the cyclic property of the trace and the fact that for a unitary representation $R^{\dagger}(h^{-1}g')=R^{\dagger}(g')R(h)$.  If instead we had applied the gauge transformation on the other charge we would get $T_h(v_3)| M_R; ( v_1, v_3 ) \rangle=| M_RR(h^{-1}); ( v_1, v_3 ) \rangle$.  In fact, these relations are true for any electric charge pair $| M_R; ( v, v' ) \rangle$ (not necessarily nearest neighbors).  

There is a connection between the action of local gauge transformations and braiding.  When a pure magnetic flux $h\in S_3$ represented as $\ket{h}$ and located at face $f$ is braided in a counterclockwise sense around a charge at location $v$, the action is,
\begin{equation}
\mathcal{R}^2_{f,v}\ket{h}\ket{M_{R};(v,v')}=\ket{h}\ket{R(h)M_R;(v,v')},
\end{equation}
where $\mathcal{R}^2_{i,j}$ is the square of the monodromy operator (it represents braiding particle at location $i$ around particle at location $j$ in a counterclockwise sense).  
If a flux with value $h$ is braided around both charges, then the action on the state is conjugation:
\begin{equation}
\mathcal{R}^2_{f,v}\mathcal{R}^2_{f,v'}\ket{h}\ket{M_{R};(v,v')}=\ket{h}\ket{R(h)M_RR(h^{-1});(v,v')},
\end{equation}
There is one state that is invariant under conjugation, the fluxless charge:
$\ket{{\bf 1}_{R};(v,v')}$.

In fact there is a simple interpretation of this state when the vertices are nearest neighbors, i.e. there is an edge $e=[v,v']$.  In that case, the pair of charges is created out of the vacuum by acting on the single spin living on the edge $e$ with
\begin{equation}
W_R(e)\ket{GS}=\ket{{\bf 1}_{R};(v,v')}
\end{equation}
where
\begin{equation}
W_R(e)=\sum_{g\in S_3} \ket{g}_e\bra{g}\chi_R^{\ast}(g)
\end{equation}
with $\chi_R$ the character of the group element in the irrep $R$. This operator is in general neither unitary nor Hermitian, but for $S_3$ it is Hermitian for all irreps.

%  When a pure magnetic flux $h\in S_3$ is braided around one of the electric charges, its acts as multiplication on the state of the charge by its representation in the irrep $R$.    This is represented by 
%\begin{equation}
%\mathcal{R}^2_{1,2}\ket{h}\ket{{\bf 1}_{R}}=\ket{h}\ket{R(h)},
%\end{equation}
%where $\mathcal{R}^2_{i,j}$ is the square of the monodromy operator (it represents braiding particle at location $i$ around particle at location $j$ in a counterclockwise sense).  Here we have adopted an ordering of the particles so that the flux is a position $1$, the electric charge at position $2$ and the anti-charge at position $3$.  The nontrivial action occurs on the joint state of the electric charge pair.  The equivalent action can be obtained by performing a local gauge transformation at the vertex location of the electric charge vertex:
%\begin{equation}
%T_h(v)\ket{{\bf 1}_{R};(v,v')}=\ket{R(h);(v,v')}.
%\end{equation}
Finally, the amplitude for the process of fusion of the electric charge pair into the vacuum is given by
\begin{equation}
F(R(h)\rightarrow vac)\equiv\bra{{\bf 1}_{R};(v,v')}R(h);(v,v')\rangle=\frac{\mbox{tr}\{R(h)\}}{|R|}.
\label{trformula}
\end{equation}

The fusion amplitudes can be measured by using an ancillary qubit to perform conditional gauge transformations and then measuring the ancilla.  For example, say we prepare 
an ancilla in $\ket{+_x}_\textrm{anc}$, then apply the controlled operation 
\begin{equation}
\ket{0}_\textrm{anc}\bra{0}\otimes {\bf 1}+\ket{1}_\textrm{anc}\bra{1}\otimes T_h(v), 
\label{controlledT}
\end{equation}
followed by measurement of the ancilla in the basis $\ket{\pm_x}$ with outcome $m=\pm 1$.  The outcome distribution satisfies:
\[
P(m=1)-P(m=-1)=\Re[\bra{{\bf 1}_{|R|};(v,v')} R(h);(v,v')\rangle],
\]
which is the real part of the fusion amplitude for $R(h)\rightarrow {\bf 1}$.  Similarly, measuring the ancilla in the basis $\ket{\pm_y}$, yields the imaginary part of the fusion amplitude for $R(h)\rightarrow$ vacuum.

Let's now calculate the expected outcomes for non-Abelian anyonic interferometry. 
Using Eqs.~(\ref{2dirrep}) and (\ref{trformula}) and the fusion amplitudes to the vacuum are:
\begin{equation}
\begin{array}{lll}
F(R_2(e)\rightarrow vac)&=&1\\
F(R_2(t_j)\rightarrow vac)&=&0\quad \forall j\\
F(R_2(c_{\pm})\rightarrow vac)&=&-\frac{1}{2}.
\end{array}
\label{amps}
\end{equation}

For the measurement proposed here it is also instructive to see how these arise
explicitly in the state overlap.  The projection operator onto $R_2$ charge is
\begin{equation}
W_{R_2}=2\ket{e}\bra{e}-\ket{c_+}\bra{c_+}-\ket{c_-}\bra{c_-}.
\end{equation}
Say we have a fluxless charge pair on nearest neighbor vertices bounding the edge $e=[v,v']$.  Then the overlap is
\begin{equation}
\begin{array}{lll}
\bra{{\bf 1}_{R_2};(v,v')}R_2(h);(v,v')\rangle &=&\bra{{\bf 1}_{R_2};(v,v')}T_h\ket{{\bf 1}_{R_2};(v,v')}\rangle \\
&=&\bra{GS}W_{R_2}(e)T_hW_{R_2}(e)\ket{W}\\
&=&\bra{GS}W_{R_2}(e)T_hW_{R_2}(e)T_h^{\dagger}\ket{W}\\
&=&\bra{GS}W_{R_2}(e)W^h_{R_2}(e)\ket{GS}\\
&=&\frac{1}{6}\mbox{tr}\{W_{R_2}(e)W^h_{R_2}(e)\}. \\
\end{array}
\end{equation}
Here we have defined the operator $W^h_{R_2}(e)=2\ket{h}_e\bra{h}-\ket{hc_+}_e\bra{hc_+}-\ket{hc_-}_e\bra{hc_-}$.  In the third line we have used the fact that the ground state is invariant under gauge transformations $T_h$ and in the last line we have used the fact that the reduced state of the any edge qudit in the ground state is maximally mixed.  Evaluating the trace we recover the fusion amplitudes in 
Eq.~(\ref{amps}) by considering the state overlap explicitly.  
For the given initial state $\ket{{\bf 1}_{R};(v,v')}$ it is also possible to infer a subset of the fusion rules for the theory without using an ancilla at all as we show in Sec.~\ref{altmeas}.

\section{Single plaquette scheme}
\label{sec:3}

\subsection{Initial state preparation}

An algorithm to prepare the ground state of $H_{\rm TO}$ in a spin lattice is given in \cite{Aguado:08a}.  However, since we only want to demonstrate the action of braiding a flux around an electric charge pair, it is sufficient to begin with the state $\ket{{\bf 1}_{R};(v,v')}$. 
Specifically, we want to create the state where the charges are on vertices $v_1$ and $v_3$ as in Fig.~\ref{fig:tri}, i.e.,
\begin{equation}
\begin{array}{lll}
\ket{{\bf 1}_{R_2};(v_1,v_3)}&=&\frac{1}{6}[2\ket{e}_2 \big( \ket{e,e}_{13}+\ket{t_0,t_0}_{13}+\ket{t_1,t_1}_{13}+\ket{t_2,t_2}_{13}+\ket{c_+,c_-}_{13}+\ket{c_-,c_+}_{13} \big) \\
&&
- \ket{c_+}_2 \big( \ket{e,c_+}_{13}+\ket{t_0,t_1}_{13}+\ket{t_1,t_2}_{13}+\ket{t_2,t_0}_{13}+\ket{c_+,e}_{13}+\ket{c_-,c_-}_{13} \big)\\
&& - \ket{c_-}_2 \big( \ket{e,c_-}_{13}+\ket{t_0,t_2}_{13}+\ket{t_1,t_0}_{13}+\ket{t_2,t_1}_{13}+\ket{c_+,c_+}_{13}+\ket{c_-,e}_{13} \big)].
\end{array}
\end{equation}
Here the subscripts indicate the qudit that this is the state for. In Fig.~\ref{fig:tri} the qudits correspond to the qutrit/qubit pairs, and the numbers are shown next to the pair. The qutrits and qubits in the pairs are further labeled by ``$a$'' for the qubits and ``$b$'' for the qutrits.

There are a number of options for how the 6-level qudit can be encoded in the qutrit/qubit pair. It is convenient to use the encoding which respects the semidirect product structure of the group $S_3$ (see Appendix~B of Ref.~\cite{Brennen:09}).  We pick for qudit~1 and~2 the encoding:
\begin{equation}
\label{eq:enc1}
\ket{e} \equiv \ket{1}_{a}\ket{0}_{b},\  \ket{t_2} \equiv \ket{0}_{a}\ket{2}_{b},\
\ket{t_0} \equiv \ket{0}_{a}\ket{0}_{b},\  \ket{c_+} \equiv \ket{1}_{a}\ket{2}_{b},\
\ket{t_1} \equiv \ket{0}_{a}\ket{1}_{b},\  \ket{c_-} \equiv \ket{1}_{a}\ket{1}_{b},
\end{equation}
and for qudit~3 the encoding:
\begin{equation}
\label{eq:enc2}
\ket{e} \equiv \ket{1}_{a}\ket{0}_{b},\  \ket{t_2} \equiv \ket{0}_{a}\ket{2}_{b},\
\ket{t_0} \equiv \ket{0}_{a}\ket{0}_{b},\  \ket{c_+} \equiv \ket{1}_{a}\ket{1}_{b},\
\ket{t_1} \equiv \ket{0}_{a}\ket{1}_{b},\  \ket{c_-} \equiv \ket{1}_{a}\ket{2}_{b}.
\end{equation}
Using this encoding, the initial state becomes
\begin{equation}
\begin{array}{lll}
\ket{{\bf 1}_{R_2};(v_1,v_3)}&=&\frac{1}{6}\ket{1}_{2a}\big(\ket{0}_{1a}\ket{0}_{3a}+\ket{1}_{1a}\ket{1}_{3a} \big) \Big[2\ket{0}_{2b} \big( \ket{0}_{1b}\ket{0}_{3b}+\ket{1}_{1b}\ket{1}_{3b}+\ket{2}_{1b}\ket{2}_{3b} \big)  \\
&&
- \ket{2}_{2b} \big( \ket{0}_{1b}\ket{1}_{3b}+\ket{1}_{1b}\ket{2}_{3b}+\ket{2}_{1b}\ket{0}_{3b} \big) - \ket{1}_{2b} \big( \ket{0}_{1b}\ket{2}_{3b}+\ket{1}_{1b}\ket{0}_{3b}\\
&&+\ket{2}_{1b}\ket{1}_{3b} \big)\Big].
\end{array}
\label{eq:encst}
\end{equation}
This factorises into an entangled pair of qubits and an entangled triple of qutrits, as indicated in Fig.~\ref{fig:tri}.

Given the initial state preparation, it remains to show how to perform controlled $T_{c_\pm}$ and $T_{t_i}$ operations.

\subsection{The controlled $T_{c_\pm}(v_1)$ operations}

The effect of the operation $T_{c_+}(v_1)$ is to exchange the basis states of qudits 1 and 2 as
\[
\ket{e} \rightarrow \ket{c_+},\   \ket{t_0} \rightarrow \ket{t_2},\ 
\ket{c_+} \rightarrow \ket{c_-},\   \ket{t_1} \rightarrow \ket{t_0},\ 
\ket{c_-} \rightarrow \ket{e},\   \ket{t_2} \rightarrow \ket{t_1}.
\]
With encoding~(\ref{eq:enc1}), this corresponds to cyclicly permuting the qutrit states as 
\[
\ket{0} \rightarrow \ket{2},\ 
\ket{1} \rightarrow \ket{0},\ 
\ket{2} \rightarrow \ket{1}.
\]
This means that, in order to measure the fusion amplitudes and perform the \emph{controlled} $T_{c_+}(v_1)$ operations of Eqn.~(\ref{controlledT}), we need to do two controlled permutations --- one on $1b$ and one on $2b$.

The operation $T_{c_-}(v_1)$ is the inverse, and gives the opposite permutation of the qutrit states
\[
\ket{0} \rightarrow \ket{1},\ 
\ket{1} \rightarrow \ket{2},\ 
\ket{2} \rightarrow \ket{0}.
\]
In this case the controlled operation is just the same, except it uses the opposite permutations.

\subsection{The controlled $T_{t_0}(v_1)$ operation}

If, instead, we want to perform the operation $T_{t_0}(v_1)$, then we need to exchange the states for qudits 1 and 2 as
\[
\ket{e} \leftrightarrow \ket{t_0},\ 
\ket{t_1} \leftrightarrow \ket{c_+},\ 
\ket{t_2} \leftrightarrow \ket{c_-}.
\]
This operation can be achieved by employing a modified \textsc{cnot} between the qubit and qutrit that only acts on two states of the qutrit. These types of modified qubit gates have already demonstrated an advantage in constructing multi-qubit gates \cite{Lanyon:08, Monz:08}. First apply such a \textsc{cnot} with the qubit as control and on states 1 and 2 of the qutrit, followed by a \textsc{not} gate on the qubit, followed by another \textsc{cnot} between the qubit and qutrit. Explicitly this sequence of operations gives
\[
\begin{array}{lll}
\ket{e}   &\equiv& \ket{1}_{a}\ket{0}_{b} \rightarrow \ket{1}_{a}\ket{0}_{b} \rightarrow \ket{0}_{a}\ket{0}_{b} \rightarrow \ket{0}_{a}\ket{0}_{b} \equiv \ket{t_0}\\
\ket{t_0} &\equiv& \ket{0}_{a}\ket{0}_{b} \rightarrow \ket{0}_{a}\ket{0}_{b} \rightarrow \ket{1}_{a}\ket{0}_{b} \rightarrow \ket{1}_{a}\ket{0}_{b} \equiv \ket{e}\\
\ket{t_1} &\equiv& \ket{0}_{a}\ket{1}_{b} \rightarrow \ket{0}_{a}\ket{1}_{b} \rightarrow \ket{1}_{a}\ket{1}_{b} \rightarrow \ket{1}_{a}\ket{2}_{b} \equiv \ket{c_+}\\
\ket{t_2} &\equiv& \ket{0}_{a}\ket{2}_{b} \rightarrow \ket{0}_{a}\ket{2}_{b} \rightarrow \ket{1}_{a}\ket{2}_{b} \rightarrow \ket{1}_{a}\ket{1}_{b} \equiv \ket{c_-}\\
\ket{c_+} &\equiv& \ket{1}_{a}\ket{2}_{b} \rightarrow \ket{1}_{a}\ket{1}_{b} \rightarrow \ket{0}_{a}\ket{1}_{b} \rightarrow \ket{0}_{a}\ket{1}_{b} \equiv \ket{t_1}\\
\ket{c_-} &\equiv& \ket{1}_{a}\ket{1}_{b} \rightarrow \ket{1}_{a}\ket{2}_{b} \rightarrow \ket{0}_{a}\ket{2}_{b} \rightarrow \ket{0}_{a}\ket{2}_{b} \equiv \ket{t_2}.
\end{array}
\]
Here the first and third arrows correspond to \textsc{cnot}s, and the second corresponds to the \textsc{not} gate on the qubit.

To achieve a controlled $T_{t_0}(v_1)$, we can take advantage of the fact that the \textsc{cnot} applied twice is the identity. Therefore, we can apply the \textsc{cnot}, then apply a \textsc{cnot} with the ancilla as control and the qubit in the qubit/qutrit pair as target, then apply another \textsc{cnot} between the qubit and qutrit. This needs to be done twice --- once for qubit/qutrit pair~1 and again for pair~2. A further simplification may be obtained by noting that the measurement of the ancilla can take place immediately after the \textsc{cnot}s with the ancilla as control. The final \textsc{cnot}s within the qubit-qutrit pairs can be omitted, as they do not affect the probabilities of the measurement results.

\subsection{The controlled $T_{t_1}(v_1)$ operation}

Next, the operation $T_{t_1}(v_1)$ exchanges the states for qudits~1 and~2 as
\[
\ket{e} \leftrightarrow \ket{t_1},\ 
\ket{t_2} \leftrightarrow \ket{c_+},\ 
\ket{t_0} \leftrightarrow \ket{c_-}.
\]
This operation can be achieved in a similar way to $T_{t_0}(v_1)$, except with \textsc{cnot}s acting on states~0 and~1 for the qutrit. Explicitly the sequence of operations gives
\[
\begin{array}{lll}
\ket{e}   &\equiv& \ket{1}_{a}\ket{0}_{b} \rightarrow \ket{1}_{a}\ket{1}_{b} \rightarrow \ket{0}_{a}\ket{1}_{b} \rightarrow \ket{0}_{a}\ket{1}_{b} \equiv \ket{t_1}\\
\ket{t_0} &\equiv& \ket{0}_{a}\ket{0}_{b} \rightarrow \ket{0}_{a}\ket{0}_{b} \rightarrow \ket{1}_{a}\ket{0}_{b} \rightarrow \ket{1}_{a}\ket{1}_{b} \equiv \ket{c_-}  \\
\ket{t_1} &\equiv& \ket{0}_{a}\ket{1}_{b} \rightarrow \ket{0}_{a}\ket{1}_{b} \rightarrow \ket{1}_{a}\ket{1}_{b} \rightarrow \ket{1}_{a}\ket{0}_{b} \equiv \ket{e}\\
\ket{t_2} &\equiv& \ket{0}_{a}\ket{2}_{b} \rightarrow \ket{0}_{a}\ket{2}_{b} \rightarrow \ket{1}_{a}\ket{2}_{b} \rightarrow \ket{1}_{a}\ket{2}_{b} \equiv \ket{c_+}\\
\ket{c_+} &\equiv& \ket{1}_{a}\ket{2}_{b} \rightarrow \ket{1}_{a}\ket{2}_{b} \rightarrow \ket{0}_{a}\ket{2}_{b} \rightarrow \ket{0}_{a}\ket{2}_{b} \equiv \ket{t_2}\\
\ket{c_-} &\equiv& \ket{1}_{a}\ket{1}_{b} \rightarrow \ket{1}_{a}\ket{0}_{b} \rightarrow \ket{0}_{a}\ket{0}_{b} \rightarrow \ket{0}_{a}\ket{0}_{b} \equiv \ket{t_0}.
\end{array}
\]

\subsection{The controlled $T_{t_2}(v_1)$ operation}

Similarly, the operation $T_{t_2}(v_1)$ exchanges the states for qudits~1 and~2 as
\[
\ket{e} \leftrightarrow \ket{t_2},\ 
\ket{t_0} \leftrightarrow \ket{c_+},\ 
\ket{t_1} \leftrightarrow \ket{c_-}.
\]
Again, the operation follows the procedure above, except with the \textsc{cnot}s acting on states~0 and~2 for the qutrit. Explicitly the sequence of operations gives
\[
\begin{array}{lll}
\ket{e}  &\equiv& \ket{1}_{a}\ket{0}_{b} \rightarrow \ket{1}_{a}\ket{2}_{b} \rightarrow \ket{0}_{a}\ket{2}_{b} \rightarrow \ket{0}_{a}\ket{2}_{b} \equiv \ket{t_2}\\
\ket{t_0} &\equiv& \ket{0}_{a}\ket{0}_{b} \rightarrow \ket{0}_{a}\ket{0}_{b} \rightarrow \ket{1}_{a}\ket{0}_{b} \rightarrow \ket{1}_{a}\ket{2}_{b} \equiv \ket{c_+}  \\
\ket{t_1} &\equiv& \ket{0}_{a}\ket{1}_{b} \rightarrow \ket{0}_{a}\ket{1}_{b} \rightarrow \ket{1}_{a}\ket{1}_{b} \rightarrow \ket{1}_{a}\ket{1}_{b} \equiv \ket{c_-}\\
\ket{t_2} &\equiv& \ket{0}_{a}\ket{2}_{b} \rightarrow \ket{0}_{a}\ket{2}_{b} \rightarrow \ket{1}_{a}\ket{2}_{b} \rightarrow \ket{1}_{a}\ket{0}_{b} \equiv \ket{e}\\
\ket{c_+} &\equiv& \ket{1}_{a}\ket{2}_{b} \rightarrow \ket{1}_{a}\ket{0}_{b} \rightarrow \ket{0}_{a}\ket{0}_{b} \rightarrow \ket{0}_{a}\ket{0}_{b} \equiv \ket{t_0}\\
\ket{c_-} &\equiv& \ket{1}_{a}\ket{1}_{b} \rightarrow \ket{1}_{a}\ket{1}_{b} \rightarrow \ket{0}_{a}\ket{1}_{b} \rightarrow \ket{0}_{a}\ket{1}_{b} \equiv \ket{t_1}.
\end{array}
\]

Other configurations of the charges turn out to be equivalent to the states above. In the case where both charges are on the lower nodes, the encoding on edges 1 and 2 is the same as before, as are the operations that need to be performed on them. The only difference is in the encoding of the third edge, which simply corresponds to a different interpretation of the experiment. The state obtained if both charges are on the right is again equivalent, though this time with different operations.  For details see~\ref{sec:equivalent}.

\section{Linear optics implementation}
\label{sec:4}
Next we consider how this scheme can be achieved using a linear optics
implementation. The core component of the linear optical demonstration is the
linear optical \textsc{cnot} of the type demonstrated in Ref.~\cite{OBrien03}. Such a \textsc{cnot} is based on dual-rail
qubits --- a single photon in a pair of optical modes. For this gate to
function correctly it is necessary to ensure that each pair of output modes
contains a single photon. This can make it problematic to chain these gates
together, as it at first appears necessary to perform quantum nondemolition
measurements of the photon number. However, it is still possible provided it
can be inferred that a single photon was present in the intermediate steps. For
example, that is the approach used in Ref.~\cite{Lanyon07}.

%\begin{figure}[!thb]
%\center{\includegraphics[width=7cm]{CNOT.eps}}
%\caption{The linear optical \textsc{cnot} of Ref.\ \cite{OBrien03}. The lines indicate the different modes, and each qubit is composed of two modes. The input qubits are $C_{\rm in}$ and $T_{\rm in}$ on the left, and the output qubits are $C_{\rm out}$ and $T_{\rm out}$ on the right. The \textsc{cnot} is performed provided one photon is detected in each of the output pairs of modes.}
%\label{fig:cnot}
%\end{figure}

To encode the qutrits we can simply extend the dual-rail encoding to a tri-rail
encoding with a single photons in three optical modes. This has the advantage
that arbitrary unitaries can be performed on these qutrits using linear optics
\cite{Reck}. 

Note that for a subset of the fusion rules, we do not need to implement the controlled~$T_j$ operations
which substantially simplifies the experiment. This process is detailed in section~\ref{altmeas}.

\subsection{Initial state preparation}

\begin{figure}[!thb]
\center{\includegraphics[scale=0.8]{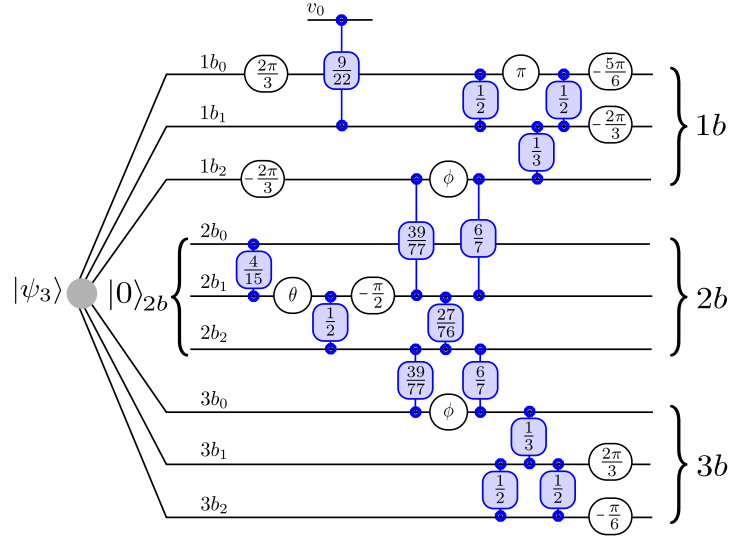}}
\caption{The scheme used to produce the desired three qutrit state from a pair
of entangled qutrits and an additional photon (the state $\ket{0}_{2b}$ corresponds to a single photon in mode $2b_0$). The horizontal lines indicate
the different optical modes, and the vertical lines indicate symmetric beam splitters connecting
modes with reflectivities shown. The open circles indicate a relative phase
shift in that mode. Note that the beam splitter connecting mode $v_0$, which is initially in a vacuum, and mode $1b_1$ is to introduce loss to balance the circuit. The phase angles $\phi$ and $\theta$, and the beam splitter convention, are 
given in the text.}
\label{fig:magic}
\end{figure}

A possible way to create the three qutrit entangled state required for $\ket{{\bf 1}_{R_2};(v_1,v_3)}$,
\begin{equation}
\begin{array}{lll}
&&\{2\ket{0}_{2b} \big( \ket{0}_{1b}\ket{0}_{3b}+\ket{1}_{1b}\ket{1}_{3b}+\ket{2}_{1b}\ket{2}_{3b} \big) - \ket{2}_{2b} \big( \ket{0}_{1b}\ket{1}_{3b}+\ket{1}_{1b}\ket{2}_{3b}+\ket{2}_{1b}\ket{0}_{3b} \big)\\
&&
- \ket{1}_{2b} \big( \ket{0}_{1b}\ket{2}_{3b}+\ket{1}_{1b}\ket{0}_{3b}+\ket{2}_{1b}\ket{1}_{3b} \big)\}/(3 \sqrt{2}),
\end{array}
\end{equation}
is to prepare the initial state $\ket{0}_{2b}\ket{\psi_{3}}_{1b,3b} $ where
\begin{equation}
\ket{\psi_3}_{1b,3b}=(\ket{0}_{1b}\ket{0}_{3b}+\ket{1}_{1b}\ket{1}_{3b}+\ket{2}_{1b}\ket{2}_{3b})/\sqrt{3}
\label{eq:qtrits}
\end{equation}
is a maximally entangled two qutrit state, and to propagate this state through
the circuit in Fig.~\ref{fig:magic}. The phase angles $\phi$ and $\theta$ in
the circuit are given by
\[
\label{eq:sol}
\theta = \arcsin \left[ \frac{10}{\sqrt{247}} \right], \;\;
\phi = \arcsin \left[\frac{7+\sqrt{3}}{2\sqrt{26}}\right] - \frac{\pi}{4}.
\]
Each of the beam splitters depicted acts in a symmetric way, transforming two
optical modes described by the boson creation operators $a^{\dagger}$ and
$b^{\dagger}$, as $a^{\dagger} \rightarrow i \sqrt{R}\;a^{\dagger} + \sqrt{1-R}\;
b^{\dagger}$ and  $b^{\dagger} \rightarrow i \sqrt{R}\;b^{\dagger} + \sqrt{1-R}\;
a^{\dagger}$, where $R$ is the reflectivity indicated.
The circuit relies on a final postselection to ensure only one photon is
present in each qutrit, which happens with a probability of $9/55$. The
remaining time, invalid qutrit states are produced such as having two photons
in three modes and these can be postselected out.

Although the circuit in Fig.~\ref{fig:magic} looks daunting, each \textsc{cnot}
gate of the type in reference~\cite{OBrien03} takes 5 beam splitter
transformations to implement, so the circuit is the same order in complexity as
three such gates. In fact, the transformation implemented by this circuit is
equivalent to a single qutrit transformation on qutrit~2 followed by a ternary
adder gate between qutrits~2 and~3 ($\ket{x,y}\rightarrow\ket{x,x\oplus y}$).
The single qubit transformation would take at least two beam splitters and the
ternary adder gate can be implemented by four qubit \textsc{cnot} gates acting
between pairs of qutrit levels. If such a naive application of \emph{qubit}
gates were used to synthesise the transformation, the circuit would already
require 22 beam splitters. The circuit in Fig.~\ref{fig:magic} has been
optimised in comparison but it may be possible to optimise it further. 

There are potentially a number of ways to prepare the entangled
state~(\ref{eq:qtrits}).  A fairly direct way, that can make use of the same
final postselection as the circuit, is to use the output of three type-I
spontaneous parametric downconversion (SPDC) crystals in a similar way to
Ref.~\cite{Lu}. Each crystal produces an infinite-dimensional entangled state
between two spatial modes,
\[
\ket{\mathrm{SPDC}} = \sqrt{1-\lambda^2} \sum_{n=0}^{\infty} 
\lambda^{n} \ket{nn}.
\]
For quantum information applications, the state is usually postselected on a
single photon being obtained in each spatial mode thus producing two photons
entangled in energy and momentum. For three independent crystals with equal
amplitudes for pair production, the state is of the form
\[
\ket{\mathrm{SPDC}}^{\otimes 3} = (1-\lambda^2)^{3/2} \sum_{n_1,n_2,n_3=0}^{\infty}
 \lambda^{n_1+n_2+n_3} \ket{n_1n_1,n_2n_2,n_3n_3}.
\]
Here the $n_j$ are the photon numbers for the modes produced by crystal $j$. If
we postselect on obtaining a single pair of photons, then since the pair may
have arisen from any of the three crystals, we obtain the
state~(\ref{eq:qtrits}) with probability $\lambda^2(1-\lambda^2)^3$. So by
using three type-I SPDC sources and an additional photon, and counting three
photons in the output, one for each qutrit, we can postselect both the
entangled input state~(\ref{eq:qtrits}) and the correct operation of the
circuit in Fig.~\ref{fig:magic}.
 
An alternative method of producing entangled qutrits has been proposed recently
\cite{herald}. That method has an advantage in that it is
heralded, but it produces qutrits encoded in the photon-number basis, which
makes it more difficult to perform subsequent operations.

\subsection{Implementing the controlled $T_{g}$ operations}

\begin{figure}[!thb]
\center{\includegraphics[width=6cm]{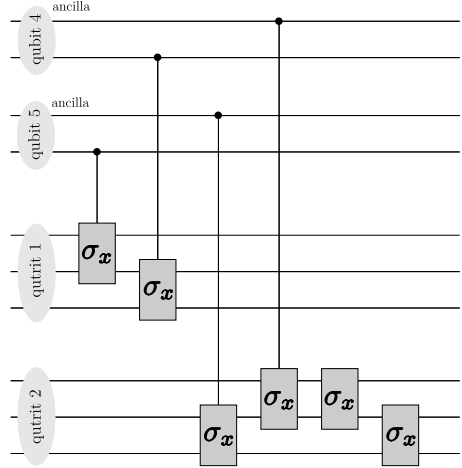}}
\caption{The sequence of linear optical \textsc{cnot}s to perform the controlled $T_{c_{+}}$ operation in the case where there is initially a single photon for each qubit and qutrit. The pairs of modes 4 and 5 are the two ancillas, and are initially in an entangled state.}
\label{fig:qubit}
\end{figure}

Given our initial state of a pair of electric charge anyons, we can measure the effect of braiding a flux around one member of the pair by performing controlled gauge transformations $T_g$ at one vertex (here vertex $v_1$) as described in Sec. \ref{sec:2}.  There are six operators $T_g$, one for each group element, and recall they involve a product of operations on spins that reside on edges that meet at vertex $v_1$.  The operator $T_e$ is trivial and we first consider how to perform the operators $T_{c_{\pm}}$ controlled on the state of a qubit ancilla.  For these two operators, it is necessary to perform controlled swaps on two qutrits in the entangled state, with a qubit as the control. Repeated linear-optical \textsc{cnot}s with the same control qubit would not work, because the linear optical \textsc{cnot} needs postselection on the final numbers of photons. One alternative is to replace the single control qubit with two control qubits, as shown in Fig.~\ref{fig:qubit}. The \textsc{cnot}s controlled by the lower rails simply give the required permutation on the first qutrit. The \textsc{cnot}s with the upper rail as control give the inverse permutation. This is followed by the desired permutation without controls, with the effect that the desired permutation is performed if the qubit photons are in the lower rails.

Provided there is a single photon in each qubit or qutrit in the input, and we postselect on single photons at the output, then the possibility of photon transfer is eliminated, and the controlled operation is performed correctly. For example, if a photon were transferred from the control (qubit 5) to the target (qutrit 1) at the first \textsc{cnot}, then a photon must be transferred from the target (qutrit 1) to the control (qubit 4) at the second to ensure that qutrit 1 has no more than one photon. However, that would imply that qubit 4 will have more than one photon.

This convenient procedure will not work if there is the possibility that there are different initial numbers of photons at the inputs. For example, if there were initially zero photons in qutrit 1, and two in qutrit 2, then a photon could be shifted from qubit 5 to qutrit 1 at the first \textsc{cnot}, then from qutrit 2 to qubit 5 at the second \textsc{cnot}. At the output, each qubit and qutrit would have a single photon, so this shift would not be detected. If the correct input state could be produced without postselection, then this procedure would work. However, for the state preparation scheme we have proposed, there is the requirement of postselection of single photons for each output qutrit.

To achieve the controlled permutation in the case where there is no guarantee of the correct photon numbers at the inputs, it would be necessary to use four ancillas in an entangled state. Each \textsc{cnot} would be applied with a different ancilla as a control. Because no ancilla is reused, any photon transfer could be detected. Although this alternative would work with the states produced by the state preparation scheme, it would be very challenging to perform experimentally due to the requirement of generating four entangled photons.

\begin{figure}[!thb]
\center{\includegraphics[width=6cm]{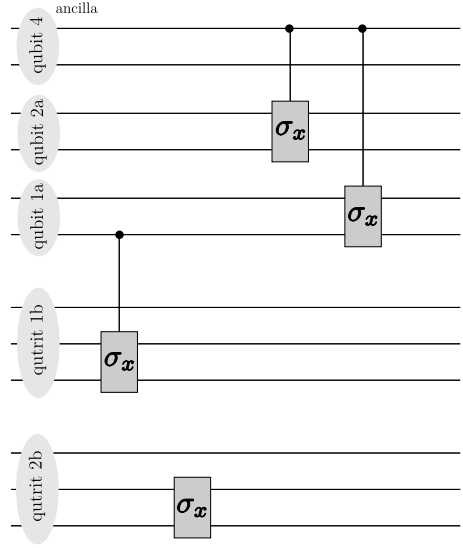}}
\caption{The sequence of linear optical \textsc{cnot}s to perform the controlled $T_{t_0}$ operation. The pair of modes 4 is the ancilla, and is initially in a superposition state. The qubit 2a is initially in the state $\ket{1}$, so the \textsc{cnot} just simplifies to a \textsc{not} operation.}
\label{fig:qubitt}
\end{figure}

To achieve the controlled $T_{t_i}$ operation, we can use the following approach. 
\begin{enumerate}
\item Generate the three qutrit entangled state.
\item Apply \textsc{cnot}s between the qubits and qutrits in pairs 1 and 2.
\item Apply \textsc{not}s on the qubits in pairs 1 and 2.
\item Apply \textsc{cnot}s with an ancilla qubit as control and the qubits in pairs 1 and 2 as targets.
\item Measure the ancilla qubit in the $\pm$ basis.
\item Perform \textsc{cnot}s between the qubits and qutrits in pairs 1 and 2 again.
\end{enumerate}
The final step 6 can be omitted, because it has no effect on the probabilities of the measurement results. In addition, the \textsc{cnot} between $2a$ and $2b$ just simplifies to a \textsc{not}, and can be performed deterministically. Steps 3 and 4 can be combined by making the \textsc{not}s controlled by the 0 state (of the ancilla), rather than 1. The scheme can then be achieved as shown in Fig.~\ref{fig:qubitt}. This example is for $T_{t_0}$; the cases of $T_{t_1}$ and $T_{t_2}$ can be addressed by applying to the \textsc{cnot} and \textsc{not} to different pairs of modes in the qutrits.

There are only three \textsc{cnot}s required in each case. There is one \textsc{cnot} between $2a$ and $2b$. Then there is a \textsc{cnot} between the ancilla qubit 4 and $1a$, and between 4 and $2a$. These three \textsc{cnot}s require that there is no photon transfer. In this case, photon transfer can be ruled out because the ancilla qubits have single photons input, and we postselect on single photons output. We also postselect on single photons at the outputs of the qutrits. This ensures that there can be no photon transfer, and the \textsc{cnot}s must have worked. This case is simpler than that for $T_{c_\pm}$, because only one ancilla qubit is required, rather than four in an entangled state, and one less \textsc{cnot} is required.

The anyonic fusion rules given above require that the probabilities for measuring the ancilla qubit in the states $+$ and $-$ are equal. It can immediately be seen that this is what will be obtained, because the \textsc{cnot} between 4 and $2a$ results in these qubits being in an entangled state. As $2a$ is then discarded without further interaction, this is equivalent to just decoherence of the ancilla qubit 4. Then the probabilitites of measuring both $+$ an $-$ are equal to 1/2. The difference in probabilities in the case of $T_{c_\pm}$ is because there is an overlap between the states.

\subsection{Measuring fusion data without controlled operations}
\label{altmeas}
An alternative, simpler, approach to extract information about the non-Abelian
statistics is to bypass controlled opeations altogether and directly measure operators $W_{R'}$ at edge $e$, in
states of the type $| M_R \vert$, where $R$, $R'$ are possibly
different irreducible representations.  These measurements give
nontrivial information about the anyon model, namely information about
a subset of the fusion rules of the theory.  In particular, $\langle
M_R \vert W_{R'} \vert M_R \rangle$ is always zero if $R^\ast \times
R$ does not contain $R'$; and by varying $M$, one can probe the
matrices $Q^{ [ R R^{\prime\ast} R^\ast ] }$ (to be defined below)
implementing the fusion rules as projectors made of $3j$ symbols.

Let us consider first the case where $R = R' = R_2$, the
two-dimensional irrep of $\mathrm{S}_3$.  In this case, we do not need
to perform controlled operations $T_{t_i}$ or $T_{c_\pm}$, and just
need the corresponding unconditional operations.  To see how this
works we compute the outcome of the measurement of $W_{R_2}$ in the gauge
transformed state $T_h(v)\ket{{\bf
   1}_{R_2};(v,v')}=\ket{R_2(h);(v,v')}$,
\begin{equation}
\begin{array}{lll}
\langle R_2(h) ;(v,v') | W_{R_2}(e) | R_2(h) ;(v,v') \rangle 
&=&\bra{{\bf 1}_{R_2};(v,v')}T^{\dagger}_hW_{R_2}(e)T_h(v)\ket{{\bf 1}_{R_2};(v,v')}\rangle \\
&=&\bra{GS}W_{R_2}(e)W^{h^{-1}}_{R_2}(e)W_{R_2}(e)\ket{GS} \\
&=&\frac{1}{6}\mbox{tr}\{(W_{R_2}(e))^2W^{h^{-1}}_{R_2}(e)\}. \\
\end{array}
\end{equation}
We find 
\begin{equation}
\langle R_2(h) ;(v,v') | W_{R_2}(e) | R_2(h) ;(v,v') \rangle =
\left\{\begin{array}{c}1 \quad h=e \\\ -\frac{1}{2} \quad h=c_{\pm} \\ 0\quad h=t_j\ \forall j\end{array}\right. .
\end{equation}
This reproduces the fusion amplitudes computed before, but this is
just an accident.  In fact, such an expectation value probes the
fusion rules of the theory:
\begin{equation}%\label{measurew}
\langle R ( h ) | W_{R'} | R ( h ) \rangle
=
\sum_{ a, \, b, \, d, \, e = 1 }^{ | R | }
\sum_{ c = 1 }^{ | R' | }
Q^{ [ R R^{\prime\ast} R^\ast ] }_{acd, \, bce}
R^\ast_{ab} (h) R_{de} (h)
\; ,
\end{equation}
where $Q^{ [ R^{(1)} R^{(2)} R^{(3)} ] }$ are the projectors onto the
vacuum fusion channel for \emph{three} irreducible representations,
that is,
\[\label{3jmatrix}
\nonumber
Q^{ [ R^{(1)} R^{(2)} R^{(3)} ] }_{ace, \, bcf}
=
\frac{ 1 }{ | G | }
\sum_g
R^{(1)}_{ab} (g)
R^{(2)}_{cd} (g)
R^{(3)}_{ef} (g)
\\
=
\sum_\alpha
q^{ [ R^{(1)} R^{(2)} R^{(3)} ], \, \alpha }_{ace}
q^{ [ R^{(1)} R^{(2)} R^{(3)} ], \, \alpha \ast }_{bcf}
\; ,
\]
where the $q^{ [ R^{(1)} R^{(2)} R^{(3)} ], \, \alpha }$ are an
orthonormal basis for the $+1$ eigenspace of $Q^{ [R^\ast
R^{\prime\ast} R] }$; they are more familiar as $3j$ symbols in the
case of angular momentum.

Measurements for $\langle R_2(h) ;(v,v') | W_{R_2}(e) | R_2(h) ;(v,v') \rangle$ can easily be performed deterministically. $T_{c_\pm}$ is just a permutation of the three modes on the qutrit. The operation $T_{t_0}$ gives the change in the encoded states
\[
\ket{1}_{a}\ket{0}_{b} \leftrightarrow \ket{0}_{a}\ket{0}_{b},\ 
\ket{0}_{a}\ket{1}_{b} \leftrightarrow \ket{1}_{a}\ket{2}_{b},\ 
\ket{0}_{a}\ket{2}_{b} \leftrightarrow \ket{1}_{a}\ket{1}_{b}.
\]
This can be achieved by a \textsc{not} gate on the qubit, and swapping states 1 and 2 for the qutrit. Then operation $T_{t_0}$ is the same, except swapping states 0 and 1 for the qutrit, and $T_{t_0}$ is the same except swapping states 0 and 2. These are all operations that can be performed deterministically. Then the operator $W_{R_2}$ can just be measured by measuring the qubits and qutrits in their computational basis.

In particular, with the operation $T_{c_+}$ the unnormalised state becomes 
\begin{equation}
\begin{array}{lll}
&&\ket{1}_{2a}\big( \ket{0}_{1a}\ket{0}_{3a}+\ket{1}_{1a}\ket{1}_{3a} \big) [2\ket{2}_{2b} \big( \ket{2}_{1b}\ket{0}_{3b}+\ket{0}_{1b}\ket{1}_{3b}+\ket{1}_{1b}\ket{2}_{3b} \big) \\
&&- \ket{1}_{2b} \big( \ket{2}_{1b}\ket{1}_{3b}+\ket{0}_{1b}\ket{2}_{3b}+\ket{1}_{1b}\ket{0}_{3b} \big)
- \ket{0}_{2b} \big( \ket{2}_{1b}\ket{2}_{3b}+\ket{0}_{1b}\ket{0}_{3b}+\ket{1}_{1b}\ket{1}_{3b} \big)].
\end{array}
\end{equation}
Then the probabilities of measuring $e$, $c_+$ and $c_-$ are $1/6$, $2/3$ and $1/6$ respectively, which gives $\langle W_{R_2} \rangle=-1/2$.

With the operation $T_{t_0}$ the unnormalised state becomes
\begin{equation}
\begin{array}{lll}
&&\ket{0}_{2a}\big( \ket{1}_{1a}\ket{0}_{3a}+\ket{0}_{1a}\ket{1}_{3a} \big)
[2\ket{0}_{2b} \big( \ket{0}_{1b}\ket{0}_{3b}+\ket{2}_{1b}\ket{1}_{3b}+\ket{1}_{1b}\ket{2}_{3b} \big) \\
&&- \ket{1}_{2b} \big( \ket{0}_{1b}\ket{1}_{3b}+\ket{2}_{1b}\ket{2}_{3b}+\ket{1}_{1b}\ket{0}_{3b} \big)
- \ket{0}_{2b} \big( \ket{0}_{1b}\ket{2}_{3b}+\ket{2}_{1b}\ket{0}_{3b}+\ket{1}_{1b}\ket{1}_{3b} \big)].
\end{array}
\label{modmeas}
\end{equation}
In this case, the state $\ket{0}_{2a}$ means that the probabilities of $e$, $c_+$ and $c_-$ are all zero, so $\langle W_{R_2} \rangle=0$.

On the other hand, for the original state the probabilities of $e$, $c_+$ and $c_-$ are $2/3$, $1/6$ and $1/6$, respectively, giving $\langle W_{R_2} \rangle=1$. This approach gives us a method to test the system deterministically, once the correct initial entangled state is prepared.

%It is something of an accident that this measurement gives us the correct fusion amplitudes.  For example, say we instead were to prepare an fluxless charge pair in the signed irrep of $R_{1}^{-}$ of $S_3$ with charge projector $W_{R_1^{-}}=\ket{e}\bra{e}+\ket{c_+}\bra{c_+}+\ket{c_-}\bra{c_-}-\ket{t_0}\bra{t_0}-\ket{t_1}\bra{t_1}-\ket{t_2}\bra{t_2}$.  Then Eq. \ref{modmeas} gives $\langle W_{R_2} \rangle=0$ always for any gauge transformation while the correction fusion amplitudes from Eq. \ref{trformula} give  $F(R_1^{-}(e)\rightarrow vac)=1\quad {\rm for}\ h=e,c_+,c_-$ and $F(R_1^{-}(e)\rightarrow vac)=-1\quad {\rm for}\ h=t_j\ \forall j$.

\section{Conclusions}
\label{sec:5}
In summary we have proposed a physical implementation of non-Abelian interferometry using entangled multi-photon states.  The realization is via a minimal construction of a non-Abelian discrete gauge theory in a single plaquette of a spin lattice.  The entire construction requires three parametric down conversion crystals to generate entangled photon pairs which are then processed using beam splitters and phase shifters to simulate the braiding and fusion of non-Abelian charges.       By choosing an initial state which is already an excited state of the model with a pair of vacuum electric charge pairs, and by judicious choice of witness to the fusion data we have found a substantially simplified protocol for the interferometry. 

Given recent advances in integrated photonic devices our protocol should be within experimental reach in the relatively near future.  Possible extensions of this model include the simulation of thermal states of topologically ordered media by introducing mixedness using ancillary degrees of freedom (see e.g. \cite{Kwiat:04}).  Such measurements could then be used to explore phenomena like thermal fragility of topological entanglement \cite{thermalfrag}.  

\section{Acknowledgements}

We would like to acknowledge helpful discussions with Prof.~A.~White.

\appendix

\section{Other charge configurations}
\label{sec:equivalent}

Alternative configurations of the charges turn out to be equivalent.
Consider the case that the charges are on the lower two nodes. The state in this case is
\begin{equation}
\begin{array}{lll}
\ket{{\bf 1}_{R_2};(v_1,v_2)}&=&\frac{1}{6}[2\ket{e}_1 \big( \ket{e,e}_{23}+\ket{t_0,t_0}_{23}+\ket{t_1,t_1}_{23}+\ket{t_2,t_2}_{23}+\ket{c_+,c_+}_{23}+\ket{c_-,c_-}_{23} \big)\\
&& - \ket{c_+}_1 \big( \ket{e,c_-}_{23}+\ket{t_0,t_1}_{23}+\ket{t_1,t_2}_{23}+\ket{t_2,t_0}_{23}+\ket{c_+,e}_{23}+\ket{c_-,c_+}_{23} \big)\\
&& - \ket{c_-}_1 \big( \ket{e,c_+}_{23}+\ket{t_0,t_2}_{23}+\ket{t_1,t_0}_{23}+\ket{t_2,t_1}_{23}+\ket{c_+,c_-}_{23}+\ket{c_-,e}_{23} \big)].
\end{array}
\end{equation}
In this case, we can use the encoding~(\ref{eq:enc1}) on all three qudits, and the state becomes the same as~(\ref{eq:encst}), except with the labels 1 and 2 interchanged.

Another case is where both charges are on the right, and the state becomes
\begin{equation}
\begin{array}{lll}
\ket{{\bf 1}_{R_2};(v_2,v_3)}&=&\frac{1}{6}[2\ket{e}_3 \big( \ket{e,e}_{12}+\ket{t_0,t_0}_{12}+\ket{t_1,t_1}_{12}+\ket{t_2,t_2}_{12}+\ket{c_+,c_+}_{12}+\ket{c_-,c_-}_{12} \big) \\
&&- \ket{c_+}_3 \big( \ket{e,c_+}_{12}+\ket{t_0,t_1}_{12}+\ket{t_1,t_2}_{12}+\ket{t_2,t_0}_{12}+\ket{c_-,e}_{12}+\ket{c_+,c_-}_{12} \big)\\
&&- \ket{c_-}_3 \big( \ket{e,c_-}_{12}+\ket{t_0,t_2}_{12}+\ket{t_1,t_0}_{12}+\ket{t_2,t_1}_{12}+\ket{c_-,c_+}_{12}+\ket{c_+,e}_{12} \big)].
\end{array}
\end{equation}
Here we would use the encoding~(\ref{eq:enc1}) on qudit 3, and the encoding~(\ref{eq:enc2}) on qubits 1 and 2. Then we would again get the state~(\ref{eq:encst}), except this time with the labels 2 and 3 interchanged.

The encodings are different on the nodes we wish to perform the operations on, but the operations are different, and the scheme is again isomorphic. To achieve the operations $T_{g}(v_2)$ we need to apply operations to qudits 1 and 3. The encoding on 3 is now the same as the encoding on 1 and 2 was previously, and we again are performing $L$ operations on this qudit, so the analysis is identical to that above. For qudit 1, we now need to perform $R$ operations, because this edge is directed inwards towards node 2. The effect of the operation $T_{c_+}(v_2)$ is to exchange the basis states of qudit 1 as
\[
\ket{e} \rightarrow \ket{c_-},\  \ket{t_0} \rightarrow \ket{t_2},\ 
\ket{c_-} \rightarrow \ket{c_+},\  \ket{t_1} \rightarrow \ket{t_0},\
\ket{c_+} \rightarrow \ket{e},\  \ket{t_2} \rightarrow \ket{t_1}. 
\]
In this case, because we are now using the encoding~(\ref{eq:enc2}) on qudit 1, this corresponds to permuting the basis states as
\[
\ket{0} \rightarrow \ket{2},\
\ket{1} \rightarrow \ket{0},\ 
\ket{2} \rightarrow \ket{1}.
\]
This is \emph{identical} to that obtained previously. The requirement to perform the $R$ operation, together with the different encoding, mean that the end result is identical.

The results are similar for the other cases. $T_{c_-}(v_2)$ is just the inverse, and the $T_{t_i}(v_1)$ are also the same as before. For example, to perform the operation $T_{t_0}(v_1)$, then we need to exchange the states for qudit 1 as
\[
\ket{e} \leftrightarrow \ket{t_0},\ 
\ket{t_1} \leftrightarrow \ket{c_-},\ 
\ket{t_2} \leftrightarrow \ket{c_+}.
\]
Again, the roles of $c_-$ and $c_+$ are exchanged, but the encoding also exchanges the roles of $c_-$ and $c_+$, so the operations required on the encoded states are identical.

Recall also that in this case, the roles of qudits 2 and 3 are exchanged for the state, so operations on qudits 1 and 3 here are equivalent to operations on qudits 1 and 2 previously. As a result of this, in each of the three possibilities for the charges, both the encoded states and the operations on these encoded states are the same. Hence any experiment can be interpreted in each of three ways: as with the charges on the nodes shown in Fig.~\ref{fig:tri}, as with the charges on the bottom two nodes, or with the charges on the right two nodes.

\end{document}